\pdfoutput=1
\documentclass[%
preprint,
 amsmath,amssymb,
 aps,
pra,
showkeys
]{revtex4-2}

\usepackage{graphicx}
\usepackage{dcolumn}
\usepackage{bm}
\usepackage{url}

\begin{document}


\title{Phase Diffusion of Light Immersed In Quantum Tides: Open Quantum System Approach}

\author{Fateme Shojaei Arani}

\author{Brahim Lamine}
\email{brahim.lamine@irap.omp.eu}

\author{Alain Blanchard}

\author{Malek Bagheri Harouni}

\affiliation{%
Department of Physics, University of Isfahan, Hezar Jerib Str., Isfahan 81746-73441, Iran;\\
}%

\affiliation{%
Université de Toulouse, UPS-OMP, IRAP, F-31400 Toulouse, France\\
}%
\affiliation{%
CNRS, IRAP, 14, avenue Edouard Belin, F-31400 Toulouse, France\\
}%
\affiliation{
Quantum Optics Group, Department of Physics, University of Isfahan, Hezar Jerib Str., Isfahan 81746-73441, Iran;\\
}%

\date{\today}

\begin{abstract}

The interaction between quantum gravitational waves (GWs) and electromagnetic (EM) fields is investigated within the open quantum system formalism, where GWs are considered as a heat bath reservoir occupying a generic state $\hat{\rho}_{\text{gws}}$. Following the quantum Langevin equations, it turns out that the correlations of the Langevin noise operator associated with the GW background directly determine the statistical properties of the EM phasor $\phi(t)$. We apply this formalism to the background of inflationary-generated primordial gravitational waves (PGW). Since this background has an astronomically large correlation time, of the order of the Hubble time $H_0^{-1}$, we show that it leads to a non-Markovian dynamics of the EM field, which causes memory effects. As a result of the Gaussianity of PGW, it turns out that the EM phasor goes through a stochastic process, which is a manifestation of the fluctuation-dissipation in EM-GW system. The variance of the EM phase smears out as $\Delta^2\varphi(t)= \Delta^2\varphi_0+ 4(t/\tau_c)^4$, where the characteristic time scale $\tau_c$ is associated with the diffusion rate caused by PGWs. The specific quartic growth of the phase noise is thus attributed to the two-mode squeezed nature of PGWs, which is inherently different from the phase diffusion induced by vacuum fluctuations of spacetime or a thermal heat bath of gravitons.

\end{abstract}

\maketitle


\section{\label{sec:1}Introduction}

The existence of a squeezed background of primordial gravitational waves (PGWs) is among the most important predictions of the inflationary models of cosmology and has yet to be verified by observation. Theoretical investigations imply that the amplifying mechanism during expanding stages of the universe has resulted in a highly squeezed spectrum of tensor perturbations, spanning from ultra-low frequency $f_{H} \sim 10^{-18}\,$ Hz up to high frequency $\sim 10^{10}\,$Hz \cite{grishchuk1977graviton, zhao2006relic}. However, in the frequency window of the current ground- and space-based gravitational wave (GW) detectors, the characteristic strain amplitude $h_c(f)$ is so small that makes the detection of PGWs challenging. Further observational efforts would then be focused on detecting the imprint of PGWs through the B-mode polarization anisotropies of the cosmic microwave background (CMB). In this regard, the LiteBIRD satellite is expected to have its results by the end of the next decade. On the other hand, a novel series of studies has grown to focus on inferring the quantum essence of inflationary-generated GWs, especially their expected squeezed nature, as discussed below. Due to the extremely weak coupling of gravity with matter, PGWs resulted from the amplification of quantum vacuum fluctuations during the Big Bang, decoupled from the rest of matter and radiation very early, and have since propagated freely throughout the universe. Hence, it is expected that they have retained their quantum properties, such as squeezing and entanglement. As a first attempt, primary investigations have examined the effect of the entanglement between opposite momenta in primordial scalar \cite{martin2016quantum} and tensor perturbations \cite{matsumura2020squeezing} on the cosmic microwave background radiation (CMB). However, significant effort has still to be made to observe such effects on the CMB.
Other promising candidates are GW detectors, such as LIGO, LISA and Einstein Telescope, that are planned to search for PGWs in their corresponding frequency windows. These detectors mainly work based on the correlations of the EM field, i.e., the phase coherence of light.
Thus, effects originating from the quantum nature of gravity on atomic- and laser-based GW detectors is being caught up in various scientific efforts \cite{abrahao2023quantum}.
In principle, these effects could be observed through matter-wave interferometry \cite{lamine2006ultimate, bassi2017gravitational}. These fluctuations can be classical, quantum, or both. 
The occurrence of decoherence is accompanied by noise injection into the system, leading to phase diffusion. 

Naively, phase diffusion refers to the random fluctuations in the phase of an optical field, arising from intrinsic noise sources such as thermal effects, spontaneous emission, and environmental perturbations. These fluctuations can degrade the coherence of light sources, leading to broader line widths and reduced stability, which are detrimental in applications requiring high phase precision \cite{feng2014quantum, ludlow2015optical, riedel2015decoherence}. In quantum optics, phase diffusion impacts the fidelity of quantum states, affecting quantum communication, cryptography, and entanglement-based technologies \cite{minavr2008phase, ren2017ground, pirandola2020advances}. Moreover, precision measurement techniques, such as interferometry and frequency comb generation, also rely on stable optical phases, where any phase instability can lead to significant measurement errors \cite{liao2024phase}. 
Thus, a central aspect of sensing in a real scenario is the interaction between the system and the environment. When one takes into account this coupling, the promised quantum enhancement is likely to be lost \cite{demkowicz2012elusive}. This has been extensively investigated in the case of a lossy interferometer \cite{datta2011quantum, demkowicz2009quantum, dorner2009optimal, kacprowicz2010experimental}. More recently, theoretical and experimental efforts have been directed at studying the limits of phase estimation in the presence of phase diffusion \cite{brivio2010experimental, genoni2011optical, genoni2012optical, escher2012quantum, knysh2013estimation}. Phase-diffusive noise reduces the visibility of interference, directly affecting precision measurements, as, for instance, in ground-based gravitational wave detection with laser interferometers \cite{tsubono1997gravitational}.

As a result, it is of crucial importance to account for underlying phase-diffusive mechanisms that affect the phase coherence of laser beams incorporated in gravitational wave detectors (GWDs) requiring incredibly high stability. The first unavoidable effect is the noise limit in the frequency stabilization due to thermal fluctuations, which is investigated in \cite{numata2004thermal}. Besides the quantum noise, the sensitivity will be mainly limited by the thermal noise of the mirror coatings \cite{gras2018direct, harry2002thermal}. Therefore, some of the proposed GWDs will be operated with cryogenically cooled silicon mirrors to lower the coating Brownian thermal noise compared to current room temperature detectors \cite{reitze2019cosmic, buikema2020sensitivity, acernese2014advanced, affeldt2014advanced}. The design and operation of stabilized laser systems applicable to GWs interferometers has been one of the main concerns of GWDs studies during the last decade and is going to be upgraded \cite{eichholz2015heterodyne}. Still, one question remains to be addressed: can GW itself defeat phase correlations of lasers? In other words, do intrinsic fluctuations of spacetime lead to loss of coherence of the laser field when it propagates through the interferometer arm lengths? 

The quantum noise suffered by two massive particles placed in a spatial superposition state is addressed in \cite{kanno2021noise}, where the Langevin equation of geodesic deviation in the presence of gravitons is obtained to evaluate the decoherence caused by squeezed GWs. The observability of these effects depends crucially on the level of squeezing of GWs. In \cite{parikh2025quantum}, Parikh et al. have proposed an experimental setup to distinguish the gravitational noise from other sources of noise with the help of nearby GW detectors. Moreover, the noise of gravitons exerted on the lengths of an interferometer was evaluated in \cite{parikh2021signatures} which could constitute direct evidence for the quantization of gravity and the existence of gravitons. In their study \cite{kanno2021indirect}, Kanno et al. consider a hypothetical situation consisting of entangled mirrors in a GW interferometer and propose the possible observation of disentanglement as a signature of squeezed PGWs. 

On the other hand, investigation of the effect of GWs on the optical field has initiated another way of exploring gravitational decoherence based on quantum optical interference tests \cite{lagouvardos2021gravitational}. It has been shown that quantum correlations in the form of Einstein-Podolski-Rosen (EPR), coded in the polarization of optical field, should survive the exposition to gravitational wave backgrounds, even at cosmological distances \cite{lamine2011large}. Besides, the influence of quantum gravity fluctuations on laser beam interferometers has been studied thoroughly \cite{abrahao2023quantum}. Especially, the effect of squeezed PGWs on the spectrum of a coherent laser beam was studied in \cite{arani2023sensing} where it is shown that the appearance of sidebands in the spectrum of light is a unique signature of the squeezed PGWs, while the location of the sidebands depends on the inflationary parameters such as the tensor-to-scalar ratio. On top of that, the loss of spatial coherence (incoherence) of distant objects caused by two-mode squeezed PGWs is investigated in \cite{arani2025revisiting} where it is shown that the van Citter-Zernike correlations are robust to the underlying noise injected by gravitons. 

The motivation for gravitational decoherence and incoherence models is multipurpose. Firstly, the existence of a fundamental decoherence process could be a solution to the quantum measurement problem, and it could explain the emergence of a classical macroscopic world \cite{reynaud2009gravitational}. Secondly, even if not observed experimentally, gravitational-induced decoherence may pave a new way to constrain the targeted gravitational background. Thirdly, it is a priority to encounter the decoherence induced by GWs in any proposed schema for GW detection, especially in phase estimation, for it would determine the ultimate bound on the phase uncertainty. The importance of this issue becomes especially remarkable in the latter case, as reducing the phase noise below the shot-noise limit plays a vital role in improving the sensitivity of interferometric detectors \cite{caves1981quantum,giovannetti2004quantum,ligo2011gravitational,motes2012phase,pezze2013ultrasensitive,tse2019quantum,genovese2021experimental}. 

In line with these motivations, we dedicate this study to present a comprehensive account of the phase diffusion and phase uncertainty induced by quantum GWs in GW interferometers. Following the Langevin-Heisenberg approach, the quantum noise operator associated with GWs is obtained, and it is shown that the EM phase correlations are directly determined by the noise correlations of the underlying GW background, averaged over the quantum state.
In order to evaluate the quantum noise injected by GWs, we apply the method of the c-number Langevin equation, which enables the treatment of quantum noise as a stochastic process for simplified statistical analysis. This method is widely used in the literature to model laser phase diffusion and line width, and in optical cavities it describes damping and input-output relations in cavity-QED \cite{gardiner1985input}. The methodology is crucial in opto-mechanics, modeling quantum backaction on mechanical oscillators \cite{aspelmeyer2014cavity} which plays a fundamental role in gravitational wave detection, describing quantum shot noise and radiation pressure noise \cite{braginsky1995quantum}. By converting operator equations into stochastic differential equations, it allows for precise calculations of decoherence, phase diffusion, and sensitivity limits in high-precision interferometric experiments. Consequently, we derive statistical properties of the EM phasor and show that it goes into a diffusion process. Ultimately, the growth of phase uncertainty induced by two-mode squeezed PGW is evaluated.

The paper is organized as follows. In Sec.~\ref{sec:2} the Heisenberg-Langevin approach is presented, and the quantum noise operator associated with GWs is found. EM phase correlation and phase diffusion are investigated in Sec.~\ref{sec:3}. Sec.~\ref{sec:5} is devoted to conclusions. 


\section{\label{sec:2} Quantum Langevin framework}

The continuum of bosonic modes of the GWs background can be thought of as an environment surrounding the propagating EM field. 
Open quantum dynamics of the EM field can be derived using the master equation or Heisenberg-Langevin approach. This would enable us to determine the ``phase diffusion" of light caused by the quantum background of GWs. Pretty similar to the quantum theory of lasers, the gravitationally induced phase diffusion is responsible for decoherence of light, which in turn leads to line-width broadening \citep{arani2023sensing}. In the following analysis, we provide a systematic description of phase diffusion of light using the quantum noise operator associated with GWs and calculate two-time correlation of the EM phase, from which we can calculate the phase uncertainty and the diffusion rate in Sec.~\ref{sec:3}.


\subsection{\label{subsec:2.A} EM Field dynamics}

\begin{figure*}[htb]
\centering
\includegraphics[
width=\textwidth]{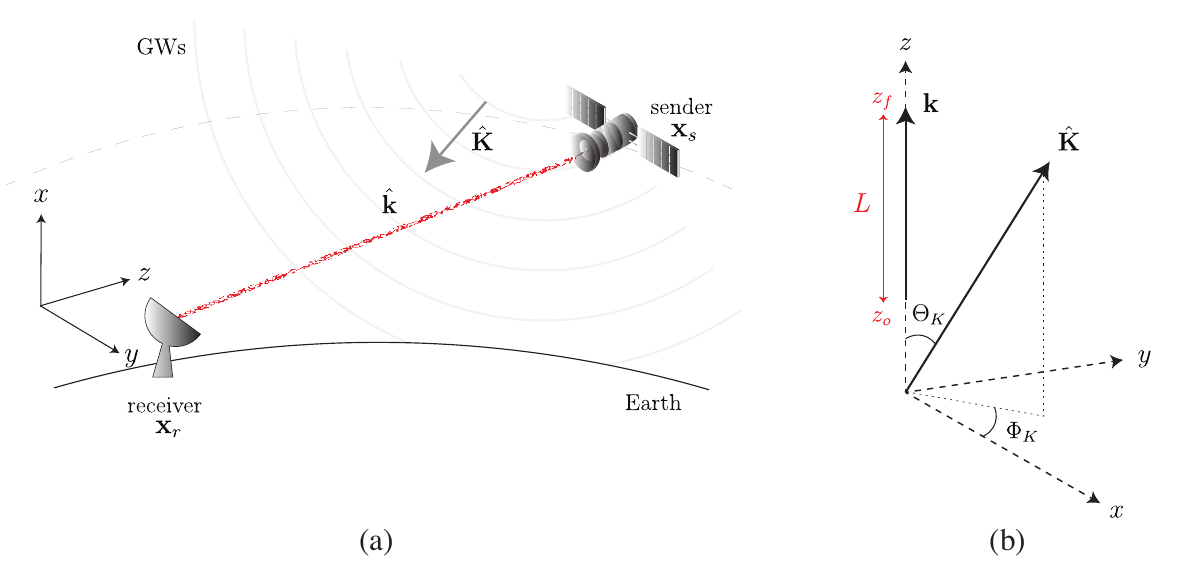}
\caption{Schematic view of the experiment. Panel (a): The coherent laser beam is sent from a sender located at $\mathbf{x}_s$, and received at a receiver at $\mathbf{x}_r$. $\hat{\mathbf{k}}$ and $\hat{\mathbf{K}}$ show the direction of propagation of the EM field and GWs with respect to some coordinate system. Panel (b): spatial extension of the EM field. It is assumed that the transverse profile of the EM field is much smaller than GW's wavelength, while the propagation length $L \equiv z_r - z_s$ can be arbitrarily large. The polar angles $(\Theta_K, \Phi_K)$ represent the direction of GW propagation with respect to $\hat{\mathbf{K}}$.}
\label{fig1}
\end{figure*}

Quantization of linearized GWs is usually done within the framework of quantum field theory (QFT), where the GW field is simply treated as a rank-$2$ tensor possessing $2$ different polarization states \cite{dirac1958theory,szczyrba1981hamiltonian, dapor2020modifications,parikh2020quantum}. Plane-wave expansion of GWs can be generally considered as
\begin{eqnarray}\label{eq:1}
\hspace*{-0.5cm}\hat{h}_{ij}(\mathbf{x},t) &=& \mathcal{C} \sum_{\gamma=+,\times} \int\frac{d^3\mathbf{K}}{(2\pi)^{3/2}} \frac{e^{\gamma}_{ij}[\hat{\mathbf{K}}]} {\sqrt{2\Omega_{K}}} \Big( \hat{b}_{K} e^{-i(\Omega t -\mathbf{K} \cdot \mathbf{x} )} \nonumber\\
&+& \hat{b}^{\dagger}_{K} e^{i(\Omega t -\mathbf{K} \cdot \mathbf{x})} \Big)\, .
\end{eqnarray}
Here, $\mathcal{C}=\sqrt{16\pi c^3} (\hbar/E_{Pl})$, ($\hat{b}_{\mathbf{K},\gamma},\hat{b}^{\dagger}_{\mathbf{K},\gamma}$) are the ladder operators of GW mode ($K \equiv \hat{\mathbf{K}},\gamma$), satisfying the bosonic commutation relation, and $\hat{e}_{ij}^{\gamma}[\hat{\mathbf{K}}]$ stands for the polarization tensor of the GW with wave vector $\mathbf{K}$ and polarization state $\gamma=+,\times$. We consider a practical situation where the laser field is sent from a sender and is received at a receiver, as shown in panel (a) of Fig.~\ref{fig1}. In panel (b) of Fig.~\ref{fig1}, we have shown the laser field propagating along the $z$ direction, starting from $z_s$ and reaching at $z_r$. Usually the cross section of the laser field is small, and one may assume that the transverse extension of the field is much smaller than the GW's wavelength $\lambda_{\text{gw}}$. Thus, we can apply the paraxial approximation and assume that the interaction with GWs is confined to a trajectory starting from $z_s$ and ending at $z_r$. Thus, the dynamical degree of the field is practically along the axial direction $z$ \cite{pang2018quantum}. However, the length of the probe $L \equiv z_r - z_s$ can be as large as the GW's wavelength, i.e. $L \sim \lambda_{\text{gw}}$. Note that, since the two-mode squeezed PGW background is homogeneous and isotropic (as in the case of vacuum or thermal fluctuations of spacetime), we have chosen the $z$ direction along the direction of propagation of the EM field. The interaction with the GW background occurs along the path $L= z_r-z_s$ where the spatial expansion of the EM field is non-vanishing. In this case, the quantum mechanical Hamiltonian describing the evolution of single-mode EM fields and GWs is encapsulated in the following Hamiltonian
\cite{arani2023sensing,abrahao2023quantum}
\begin{eqnarray}\label{eq:2}
\hat{H} = \hat{H}_{\text{em}}^{(0)} + \hat{H}_{\text{gw}}^{(0)} +  \hat{H}_{\text{int}} \, ,
\end{eqnarray}
where 
\begin{eqnarray}\label{eq:3}
\hat{H}_{\text{em}}^{(0)} &=& \hbar \omega \hat{a}^{\dagger}\hat{a}\, ,\\
\hat{H}_{\text{gw}}^{(0)} &=& \sum _{\gamma=+,\times}\int d^3\mathbf{K} \, \hbar\,\Omega_K \, \hat{b}^{\dagger}_{\mathbf{K},\gamma} \hat{b}_{\mathbf{K},\gamma} \, , \nonumber
\end{eqnarray}
describe free evolution of EM and GWs, respectively, and the interaction between a single-mode plane EM field passing by the continuum of GWs is described by
\begin{eqnarray}\label{eq:4}
\frac{\hat{H}_{\text{int}}}{\hbar} &=& -\frac{1}{2} \frac{\sqrt{16\pi c^3}}{(2\pi)^{3/2}} \Big( \frac{\hbar\omega}{E_{Pl}} \Big) \sum_{\gamma=+,\times} \int \frac{d^3\mathbf{K}}{\sqrt{2\Omega_K}} F_{\gamma} (\hat{\mathbf{K}},\hat{\mathbf{k}}) \nonumber \\
&& \Big( \hat{b}_{\mathbf{K},\gamma}\, g_{\mathbf{K}}(z_s, z_r) + \hat{b}_{\mathbf{K},\gamma}^{\dagger} \, g^{\ast}_{\mathbf{K}}(z_s, z_r) \Big) \hat{a}^{\dagger} \hat{a}\, ,
\end{eqnarray}
where the Hamiltonian is written in the Schr\"{o}dinger picture. Here, ($\hat{a},\hat{a}^{\dagger}$) represents ladder operators of the EM field of mode $k\equiv(\mathbf{k},\sigma)$, with $\mathbf{k}$ and $\sigma$ being its wave vector and polarization, respectively. The function
\begin{eqnarray}\label{eq:5}
F_{\gamma}(\hat{\mathbf{K}},\hat{\mathbf{k}})=\hat{e}_{ij}^{\gamma}[\hat{\mathbf{K}}]\hat{\mathbf{k}}_{i}\, \hat{\mathbf{k}}_j \, ,
\end{eqnarray}
is called the detector pattern function and bears the configuration of the propagation direction of the EM field $\hat{\mathbf{k}}$ with respect to the incoming GWs polarization. In the traceless-transverse (TT) gauge, the two independent polarization states $\gamma = +, \times$ can be expressed in terms of two unit vectors $\hat{\mathbf{n}}$ and $\hat{\mathbf{m}}$ orthogonal to the propagation direction $\hat{\mathbf{K}}$ and to each other, and the polarization tensor is characterized by
\begin{eqnarray}\label{eq:6}
e_{ij}^{+}[\hat{\mathbf{K}}] &=& \hat{n}_i\hat{n}_j - \hat{m}_i \hat{m}_j \, ,\\
e_{ij}^{\times}[\hat{\mathbf{K}}] &=& \hat{n}_i\hat{m}_j + \hat{n}_j \hat{m}_i \, .\nonumber
\end{eqnarray}
For instance, the detector pattern function $F_{\gamma}(\mathbf{K},\mathbf{k})$ vanishes in the special case of parallel propagation $\hat{\mathbf{k}} \parallel \hat{\mathbf{K}}$. The function $g_{\mathbf{K}}(z_s, z_r)$ bears the physical size and location of the interferometric probe in the space and is a result of variation of gravitational waves over spatial extent of the probe \cite{pang2018quantum}, and is defined by
\begin{eqnarray}\label{eq:7}
g_{\mathbf{K}}(z_s, z_r) &\equiv& \frac{1}{L} \int_{z_s}^{z_r} dz \, e^{i K_z z} \\
&=& e^{iK_z z_s} \frac{ e^{iK_z L} - 1 }{iK_z L}\,, \nonumber
\end{eqnarray}
where $L = z_r-z_s$. In the above equation, $K_z$ stands for the $z$-component of the GW wave vector $\mathbf{K}$.
In the limit that the length of the detector is small compared to GW's wavelength $K_z L \ll 1$, one obtains $g_{\mathbf{K}}(z_s,z_r) \rightarrow e^{iK_z z_s}$. As we are going to see below, physical quantities appear with $|g_{\mathbf{K}}(z_s,z_r)|^2$ in the presence of homogeneous GWs such as vacuum, thermal, and two-mode squeezed PGWs, that recast to unity in the small detector approximation.
In the following, we retain spatial dependence in all equations and apply the small-detector approximation when evaluating final results.

Hamiltonian Eqs.~(\ref{eq:2}-\ref{eq:4}) determines the Heisenberg equations of motion governing the field operators according to
\begin{eqnarray}\label{eq:8}
\dot{\hat{a}}(t) &=& \frac{i}{\hbar} [\hat{H},\hat{a}(t)]\, ,\\
\dot{\hat{b}}_{\mathbf{K},\gamma}(t) &=& \frac{i}{\hbar} [\hat{H},\hat{b}_{\mathbf{K},\gamma}(t)]\, ,\nonumber
\end{eqnarray}
It can be seen that dynamics of gravitons induced by EM-GWs interaction is determined by
\begin{eqnarray}\label{eq:9}
\hat{b}_{\mathbf{K},\gamma}(z_s, z_r,t) &=& \hat{b}_{\mathbf{K},\gamma}\, e^{-i\Omega t} \\
&+& \left( \frac{i}{2} \frac{\sqrt{16\pi c^3}}{(2\pi)^{3/2}} \frac{\hbar\omega}{E_{Pl}} \frac{1}{\sqrt{2\Omega_K}} F_{\gamma}(\hat{\mathbf{K}},\hat{\mathbf{k}})\right) \nonumber \\
&\times & g^{\ast}_{\mathbf{K}}(z_s, z_r) \int_{0}^{t} \hat{a}^{\dagger}(t') \, \hat{a}(t') e^{-i\Omega(t-t')} dt'\, .\nonumber
\end{eqnarray}
However, the back-action of the second term on the EM field is of the order of $\big(\frac{\hbar\omega}{E_{Pl}}\big)^2$ and may be disregarded in the first-order approximation. Putting Eq.~(\ref{eq:9}) back into the Hamiltonian Eq.~(\ref{eq:4}), the resulting equation of motion for the EM operators comes out as follows,
\begin{eqnarray}\label{eq:10}
\dot{\hat{a}}(z_s, z_r,t) = -i\omega \, \hat{a}+ i \hat{f}(z_s, z_r,t)\, \hat{a}\, ,
\end{eqnarray}
where
\begin{eqnarray}\label{eq:11}
\hspace*{-0.5cm}\hat{f}(z_s, z_r,t) &\equiv& \frac{1}{2}\frac{\sqrt{16\pi c^3}}{(2\pi)^{3/2}} \left( \frac{\hbar\omega}{E_{\text{Pl}}}\right) \sum_{\gamma}\int \frac{d^3\mathbf{K}}{\sqrt{2\Omega_K}} F_{\gamma}(\hat{\mathbf{K}},\hat{\mathbf{k}}) \\
&\times &\big( \hat{b}_{\mathbf{K},\gamma} g_{\mathbf{K}}(z_s, z_r) \, e^{-i\Omega t} + \hat{b}^{\dagger}_{\mathbf{K},\gamma} g^{\ast}_{\mathbf{K}}(z_s, z_r)\, e^{i\Omega t} \big)\, ,\nonumber
\end{eqnarray}
is called the noise operator since it depends only on the reservoir operators $(\hat{b}_{\mathbf{K},\gamma},\hat{b}^{\dagger}_{\mathbf{K},\gamma})$. Eq.~(\ref{eq:10}) implies that the GW background acts as a quantum noise source. Indeed, by combining Eqs.~(\ref{eq:1}) and (\ref{eq:11}), it can be seen that the quantum noise operator in the small detector limit $KL \ll 1$ is proportional to the strain field $h_{ij}(t)$ through
\begin{eqnarray}\label{eq:12}
\hat{f}(z_s,z_r,t) = \frac{\hbar \omega}{2} \hat{h}_{ij}(z_s,z_r,t) \, \hat{\mathbf{k}}_i \, \hat{\mathbf{k}}_j \,,
\end{eqnarray}
thus the noise correlations are directly determined by quantum correlations of GWs.
One may remove the fast frequency dependence of $\hat{a}(z_s, z_r,t)$ by transforming to the slowly varying annihilation operator, $\hat{\tilde{a}}(z_s, z_r,t) \equiv \hat{a} (z_s, z_r,t) e^{i\omega t}$ which results in the following equation for the new field operator:
\begin{eqnarray}\label{eq:13}
\dot{\hat{\tilde{a}}}(z_s, z_r,t) = i \hat{f}(z_s, z_r,t) \, \hat{\tilde{a}}(z_s, z_r,t) \,.
\end{eqnarray}
By integrating over time, one may find the behavior of the EM field operator,
\begin{eqnarray}\label{eq:14}
\hat{\tilde{a}}(z_s, z_r,t) = \int_{0}^{t} \hat{f}(z_s,z_r,t') \hat{\tilde{a}}(z_s,z_r,t') dt' + \hat{\tilde{a}}(z_s,z_r,0)\, . \qquad
\end{eqnarray}
In general, the time development of the field depends on its past. If the GW background is delta-correlated, the EM field dynamics is Markovian and there is no memory effect. However, as we shall see in the following, PGWs background possesses a long correlation time, thus is highly-correlated, which results in non-Markovian dynamics for the EM field.


\subsection{\label{subsec:2.C}Langevin noise operator correlations}

In order to show the physical origin of phase diffusion, it is convenient to work with polar coordinates $|\alpha|$ and $\phi$ which are defined by $\alpha = |\alpha| e^{i\phi}$, where $\alpha$ is a c-number variable corresponding to the mean field $\langle \hat{a} \rangle = \alpha$. In this way, one solves the c-number equation corresponding to the operator-value equation Eq.~(\ref{eq:13}). This correspondence is realized under specific circumstances as follows.

Basically, the interaction Hamiltonian Eq.~(\ref{eq:4}) is nonlinear in the field operators (intensity-dependent coupling$\hat{a}^{\dagger} \hat{a}$) and resembles the cavity-optomechanical interaction. The standard method to solve such a system is based on a straightforward linearization in which the cavity field is split into an average coherent amplitude and a fluctuating term, according to $\hat{a} = \alpha + \delta\hat{a}$ \cite{aspelmeyer2014cavity}. This mean field description is valid as long as the amplitude of the field is large, in the sense that $|\langle \hat{a} \rangle | \gg |\delta\hat{a}|$. The time development of fluctuations becomes important when the coupling of the system to its environment is strong enough, or fluctuation-initiated effects such as cooling, squeezing, and stability of the system matter \cite{aspelmeyer2014cavity}. In the case of GWs, the coupling to the EM field is extremely weak (proportional to $\frac{\hbar\omega}{E_{\text{Pl}}}$). Moreover, in a typical GW interferometer, the EM field occupies a large number of photons so that $|\langle \hat{a} \rangle | \gg |\delta\hat{a}|$ and a semiclassical treatment can be adopted safely. Thus, we may proceed by neglecting quantum fluctuations of the EM field and consider the time development of the mean values. On the other hand, the PGW background is placed in a two-mode squeezed state, which is Gaussian, and the quantum noise operator $\hat{f}(z_s,z_r,t)$ is fully characterized by its two-point correlators. This means we can replace it with a classical stochastic process $\mathcal{F}(z_s,z_r,t)$.

Hence, this method facilitates the computation of the phase diffusion and lets us determine statistical properties of the EM phasor, pretty similar to the well-established analysis of the EM phase in quantum optics and laser physics. The Langevin operator equation, Eq.~(\ref{eq:13}) then corresponds to the following c-number equation, 
\begin{eqnarray}\label{eq:15}
\dot{\alpha}(z_s, z_r,t) = i \mathcal{F}(z_s, z_r,t) \, \alpha(z_s, z_r,t) \,,
\end{eqnarray}
where $\mathcal{F}(z_s, z_r,t)$ is now a classical noise function that possesses the same correlations as the quantum noise operator $\hat{f}(z_s, z_r,t)$ defined by Eq.~(\ref{eq:11}), such that 
\begin{eqnarray}\label{eq:16}
\big\langle \mathcal{F}(z_s, z_r,t) \big \rangle &\equiv& \big\langle \hat{f}(z_s, z_r,t) \big\rangle_{\hat{\rho}_{\text{gw}}} \, , \\
\big\langle \mathcal{F}(z_s, z_r,t) \mathcal{F}(z_s, z_r,t') \big \rangle &\equiv& \big\langle \hat{f}(z_s, z_r,t) \hat{f}(z_s, z_r,t') \big\rangle_{\hat{\rho}_{\text{gw}}} \, . \nonumber
\end{eqnarray}
In fact, the classical noise function $\mathcal{F}(z_s,z_r,t)$ is equal to the trace of the symmetrized product of the quantum noise operators, and the anti-symmetrized product contains quantum effects that are neglected here. It is shown that the latter is negligible whenever the temperature of the PGW background is very high and the EM-GW coupling is very small, with the product of both being finite \cite{lamine:tel-00006936}. 
From Eq.~(\ref{eq:10}), it is straightforward to check that the mean number of photons is conserved
\begin{eqnarray}\label{eq:17}
\frac{d}{dt} \big( \hat{a}^{\dagger} \hat{a} \big) = 0 \, ,
\end{eqnarray}
Thus, one has $|\alpha|^2 = \bar{n} = cte$ and Eq.~(\ref{eq:15}) implies that the phase of the EM field obeys following equation
\begin{eqnarray}\label{eq:18}
\dot{\phi}(z_s, z_r,t) = \mathcal{F}(z_s, z_r,t)\, .
\end{eqnarray}
Hence, the rate of change of phase is sourced by the Langevin force, $\mathcal{F}(z_s, z_r,t)$ which is a stochastic noise. The phase noise correlations are closely related to correlations of GWs and are specified by

\begin{eqnarray}\label{eq:19}
\big\langle \mathcal{F}(z_s, z_r,t) \big\rangle &=& \big\langle \hat{f}(z_s, z_r,t) \big\rangle _{\hat{\rho}_{\text{gw}}} \\
&=& \left( \frac{1}{2} \frac{\sqrt{16\pi c^3}}{(2\pi)^{3/2}} \frac{\hbar\omega}{E_{Pl}} \right) \sum_{\gamma}\int\frac{d^3\mathbf{K}}{\sqrt{2\Omega_{K}}} F_{\gamma}(\hat{\mathbf{K}},\hat{\mathbf{k}}) \Big\langle \hat{b}_{\mathbf{K},\gamma}\, g_{\mathbf{K}}(z_s, z_r) \, e^{-i\Omega_K t} + \, c.c. \Big \rangle_{\hat{\rho}_{\text{gw}}}, \nonumber\\
\hspace*{-1cm}\big\langle \mathcal{F}(z_s, z_r,t) \mathcal{F}(z_s, z_r,t') \big\rangle &\equiv& \big\langle \hat{f}(z_s, z_r,t) \hat{f}(z_s, z_r,t') \big\rangle_{\hat{\rho}_{\text{gw}}} \nonumber \\
&=& \left(\frac{1}{2} \frac{\sqrt{16\pi c^3}}{(2\pi)^{3/2}} \frac{\hbar\omega}{E_{Pl}} \right)^2 \sum_{\gamma,\gamma'}\int\frac{d^3\mathbf{K} d^3\mathbf{K}'}{\sqrt{4\Omega_{K} \Omega_{K'}}} F_{\gamma}(\hat{\mathbf{K}},\hat{\mathbf{k}}) F_{\gamma'}(\hat{\mathbf{K}}',\hat{\mathbf{k}}) \nonumber \\
&\times& \bigg( \Big\langle \hat{b}_{\mathbf{K},\gamma} \hat{b}_{\mathbf{K}',\gamma'} \Big \rangle_{\hat{\rho}_{\text{gw}}} \, g_{\mathbf{K}}(z_s, z_r) \, g_{\mathbf{K}'}(z_s, z_r) e^{-i(\Omega_K t + \Omega_{K'} t')} \nonumber\\
&+& \Big\langle \hat{b}_{\mathbf{K},\gamma} \hat{b}^{\dagger}_{\mathbf{K}',\gamma'} \Big\rangle_{\hat{\rho}_{\text{gw}}} \, g_{\mathbf{K}}(z_s, z_r) \, g^{\ast}_{\mathbf{K}'}(z_s, z_r) \,  e^{-i(\Omega_K t - \Omega_{K'} t')} \nonumber\\
&+& \Big\langle \hat{b}^{\dagger}_{\mathbf{K},\gamma} \hat{b}_{\mathbf{K}',\gamma'} \Big\rangle_{\hat{\rho}_{\text{gws}}} \, g^{\ast}_{\mathbf{K}}(z_s, z_r) \, g_{\mathbf{K}'}(z_s, z_r) \,  e^{i(\Omega_K t - \Omega_{K'} t')} \nonumber\\
&+& \Big\langle \hat{b}^{\dagger}_{\mathbf{K},\gamma} \hat{b}^{\dagger}_{\mathbf{K}',\gamma'} \Big\rangle_{\hat{\rho}_{\text{gw}}} \, g^{\ast}_{\mathbf{K}}(z_s, z_r) \, g^{\ast}_{\mathbf{K}'}(z_s, z_r)\, e^{i(\Omega_K t + \Omega_{K'} t')} \bigg) \, .\nonumber
\end{eqnarray}

Thus, statistical features of Langevin force drastically depend on the quantum state of GWs, coded in the density matrix $\hat{\rho}_{\text{gw}}$. Eq.~(\ref{eq:19}) explicitly shows the contribution of the mean number of gravitons $\langle \hat{b}^{\dagger}_{\mathbf{K},\gamma} \hat{b}_{\mathbf{K},\gamma} \rangle$, as well as the contribution of quantum correlations of type $\langle \hat{b}_{\mathbf{K},\gamma} \hat{b}_{\mathbf{K}',\gamma'} \rangle$ in phase diffusion. Moreover, Eq.~(\ref{eq:19}) shows time correlations of the PGW background, which is completely different from the form of a Dirac delta function. The mean value of the Langevin force is vanishing if GWs are placed in vacuum, two-mode squeezed, or thermal states, 
\begin{eqnarray}\label{eq:20}
\big\langle \mathcal{F}(z_s, z_r,t) \big\rangle &=& 0 \, ,
\end{eqnarray} 
Thus, the stochastic Langevin force of PGWs has zero mean value. On the other hand, quantum correlations in two-mode squeezed PGWs are specified by
\begin{eqnarray}\label{eq:21}
\hspace*{-0.25cm}\langle \hat{b}_{\mathbf{K},\gamma} \hat{b}_{\mathbf{K}',\gamma'} \rangle_{\text{ts}} &=& - \, e^{2i\phi_K(\eta_H)} \sinh r_K(\eta_H) \cosh r_K(\eta_H) \nonumber \\
&\times& \delta_{\gamma\gamma'} \delta^{(3)}(\mathbf{K}+\mathbf{K}') \, , \nonumber \\
\hspace*{-0.25cm}\langle \hat{b}^{\dagger}_{\mathbf{K},\gamma} \hat{b}^{\dagger}_ {\mathbf{K}',\gamma'}\rangle_{\text{ts}} &=& - \, e^{-2i\phi_K(\eta_H)} \sinh r_{K}(\eta_H) \cosh r_{K}(\eta_H) \nonumber\\
&\times & \delta_{\gamma\gamma'} \delta^{(3)}(\mathbf{K}+\mathbf{K}') \nonumber \, , \\
\hspace*{-0.25cm}\langle \hat{b}_{\mathbf{K},\gamma} \hat{b}^{\dagger}_{\mathbf{K}',\gamma'} \rangle_{\text{ts}} & = & \cosh^2 r_{K}(\eta_H) \delta_{\gamma\gamma'} \delta^{(3)}(\mathbf{K}-\mathbf{K}') \, ,\nonumber \\
\hspace*{-0.25cm}\langle \hat{b}^{\dagger}_{\mathbf{K},\gamma} \hat{b}_{\mathbf{K}',\gamma'} \rangle_{\text{ts}} &=& \sinh^2 r_K(\eta_H) \delta_{\gamma\gamma'} \delta^{(3)}(\mathbf{K} - \mathbf{K}') \, . 
\end{eqnarray}
In these equations, $(r_K(\eta_H)),\Phi_K(\eta_H))$ represent the present-day values of the squeezing amplitude and phase of PGWs, and $\eta_H$ denotes the present conformal time. Due to the linear momentum conservation, two modes with opposite momenta $\mathbf{K}$ and $-\mathbf{K}$ are highly-entangled. Technically, the main quantity that will be measured by a typical GW interferometer is the power spectrum of PGWs, obtained from the spectral amplitude $h(K,\eta_H)$ according to $P(K,\eta_H) = h^2(K,\eta_H)$. The spectral amplitude of PGWs depends on various parameters, namely, the cosmological model and successive expansionary stages of the universe. The duration of the long-wavelength regime, where the mode is outside of the Hubble horizon, determines how much the mode is amplified through the particle creation mechanism. Investigation of $h(K,\eta_H)$ has been the subject of various studies \cite{tong2012revisit,tong2013using}. Accordingly, the squeezing amplitude $r_K$ is related to the present spectral amplitude through $e^{r_K}(\eta_H) = \frac{1}{8\sqrt{\pi}} \frac{\ell_{H}}{\ell_{\text{Pl}}} \left(\frac{K_{H}}{K}\right) h(K,\eta_H)$ \cite{arani2025revisiting,grishchuk2001relic}. Hence, the graviton content in a given mode $K$ is determined by $\bar{n}_{\text{ts}} = \sinh^2 r_K \simeq e^{2r_K}/4$.
By plugging Eq.~(\ref{eq:21}) into Eq.~(\ref{eq:19}) and performing straightforward calculations, it turns out that two-time noise correlations are determined by
\begin{eqnarray}\label{eq:22}
  \big\langle \mathcal{F}(z_s, z_r,t) \mathcal{F}(z_s, z_r,t') \big\rangle &=& \frac{c^2}{2\pi^2} \left(\frac{\hbar\omega}{E_{Pl}} \right)^2 \sum_{\gamma=+,\times} \int d(\cos\Theta_K) d\Phi_K F^2_{\gamma}(\hat{\mathbf{K}},\hat{\mathbf{k}}) \int_{K_E}^{K_1} K dK \, \vert g_{\mathbf{K}}(z_s, z_r) \vert^2 \nonumber\\
&&\bigg\{ \cosh^2r_K(\eta_H) \, e^{-i\Omega_K(t-t')}+\sinh^2 r_K(\eta_H) \, e^{i\Omega_K (t-t')} \nonumber\\
  &&- \sinh 2r_K(\eta_H) \, \Re\Big[ e^{i\big(2\phi_K(\eta_H) - \Omega_K (t+t') \big)} \Big] \bigg\} \,. 
\end{eqnarray}
Here, $d(\cos\Theta_K) d\Phi_K$ represents the solid angle element of the wave vector $\mathbf{K}$ (see Fig.~\ref{fig1}). Moreover, the integral limits are determined by $K_E$ and $K_1$ corresponding to the lowest and highest frequencies of the PGWs. Each term on the r.h.s of Eq.~(\ref{eq:25}) shows the Fourier transform of the PGW's spectrum; hence, the two-time correlation of the Langevin noise is nothing but the Fourier transform of the PGW's spectrum, which will appear as a memory kernel in specific observables of the EM field (see Eq.~(\ref{eq:25})). This is a general result in the context of open quantum systems \cite{vacchini2010exact,yang2013master}. In case of large squeezing amplitude $r_K(\eta_H) \gg 1$ and vanishing squeezing phase $\Phi_K(\eta_H) =0 $ for super-horizon modes, Eq.~(\ref{eq:22}) simplifies to the following form:
\begin{eqnarray}\label{eq:23}
\big\langle \mathcal{F}(L,t) \mathcal{F}(L,t') \big\rangle &=& \frac{1}{2\pi^2} \left(\frac{\hbar\omega}{E_{Pl}} \right)^2 \sum_{\gamma=+,\times} \int d(\cos\Theta_K) \, d\Phi_K F^2_{\gamma}(\hat{\mathbf{K}},\hat{\mathbf{k}}) \int_{\Omega_E}^{\Omega_1} d\Omega_K \Omega_K \, \Big( \frac{\sin(\Omega_K L\cos\Theta_K/c)}{\Omega_K L\cos\Theta_K/c} \Big)^2 \nonumber\\
&\times& \bigg\{ e^{-i\Omega_K(t-t')} + 2 \, \bar{n}_{ts}\, \Big( \cos(\Omega_K(t-t')) - \cos(\Omega_K(t+t') \Big) \bigg\} \,. 
\end{eqnarray}
where we have used definition Eq.~(\ref{eq:7}) and replaced $|g_{\mathbf{K}}(L)|^2 = (\frac{\sin(KL\cos\Theta_K)}{KL\cos\Theta_K})^2$. In the small detector approximation, $KL \ll 1$ which is fulfilled for the major frequency part of PGW, one has $|g_{\mathbf{K}}(L)|^2 \rightarrow 1$. The first term in Eq.~(\ref{eq:23}) shows the contribution of vacuum fluctuations of GWs, while the second term, which is proportional to the mean number of gravitons in the two-mode squeezed state $\bar{n}_{\text{ts}}$, shows the contribution of PGWs in the noise correlation function.
This leads to the fact that the EM dynamics is non-Markovian due to the large correlation time of the PGWs background. The memory effects and non-Markovian dynamics are more evident in the following, where we investigate phase diffusion due to the interaction with a bath of gravitons in different quantum states, namely, vacuum fluctuations, two-mode squeezed, and thermal states.


\section{\label{sec:3} GWs - induced phase diffusion}

\subsection{\label{subsec:3.A} Phase uncertainty}

By integrating Eq.~(\ref{eq:18}) from $t=0$ to a given instant of time $t$, one obtains
\begin{eqnarray}\label{eq:24}
\phi(z_s, z_r,t) = \int_{0}^{t} \mathcal{F}(z_s, z_r,t') \, dt' + \phi_0 \, .
\end{eqnarray}
In the above equation, $\phi_0$ shows the initial value of the EM phase. Statistical properties of $\phi(z_s, z_r,t)$ can be found with the help of the one-point and two-point correlations, namely,
\begin{eqnarray}\label{eq:25}
&&\big\langle \phi(z_s, z_r,t) \big\rangle = \int_{0}^{t} \big\langle \mathcal{F}(z_s, z_r,t') \big\rangle \, dt' + \big \langle \phi_0 \big\rangle \, , \quad\\
&& \big\langle \phi(z_s, z_r,t) \, \phi(z_s, z_r,t') \big\rangle = \int_{0}^{t} dt'' \int_{0}^{t'} dt''' \nonumber\\
&& \big\langle \mathcal{F}(z_s, z_r,t'') \, \mathcal{F}(z_s, z_r,t''') \big\rangle \nonumber\\
&+& \big\langle\phi_0\big\rangle \Big( \int_{0}^t dt'' \big\langle \mathcal{F}(z_s, z_r,t'') \big\rangle + \int_{0}^{t'} dt'' \big\langle \mathcal{F}(z_s, z_r,t'') \big\rangle \Big) \nonumber\\
&+& \big\langle \phi_0 \big\rangle^2 \nonumber \, .
\end{eqnarray}
In particular, phase diffusion can be addressed by considering autocorrelations $\langle \phi^2(z_s, z_r,t) \rangle$. We now focus on the time dependence of this autocorrelation. With the help of Eqs.~(\ref{eq:20}, \ref{eq:25}), the phase variance $\Delta^2\phi(z_s, z_r,t)$ is determined by 
\begin{eqnarray}\label{eq:26}
\hspace*{-0.25cm} \Delta^2\phi(z_s, z_r,t) &\equiv& \big\langle \phi^2(z_s, z_r,t) \big\rangle - \big\langle \phi(z_s, z_r,t) \big\rangle^2 \\
&=& \Delta^2\phi_0 + \nonumber\\
&&\int_{0}^{t} dt' \int_{0}^{t} dt'' \big\langle \mathcal{F}(z_s, z_r,t') \mathcal{F}(z_s, z_r,t'') \big\rangle \, . \nonumber
\end{eqnarray}
Eq.~(\ref{eq:26}) explicitly shows that statistical properties of the stochastic function $\mathcal{F}(z_s, z_r,t)$ determine the EM phase noise. In particular, PGWs background leads to non-Markovian dynamics due to its long correlation time of the order of the Hubble time $t_H \sim \Omega_H^{-1}$. 


\subsection{\label{subsec:3.B} Phase uncertainty in the presence of two-mode squeezed PGWs}

With the help of Eq.~(\ref{eq:23}), it is straightforward to calculate the contribution of two-mode squeezed PGWs in phase uncertainty Eq.~(\ref{eq:26}), which turns out to be given by 
\begin{eqnarray}\label{eq:27}
\Delta^2\phi(t) &=& \Delta^2\phi_0 + \mathcal{N}_{pgw}(t) \, ,
\end{eqnarray}
where the phase noise induced by PGWs is defined by
\begin{eqnarray}\label{eq:28}
\mathcal{N}_{pgw} &=& \frac{1}{\pi^2} \left(\frac{\hbar\omega}{E_{Pl}} \right)^2 \sum_{\gamma=+,\times} \int d(\cos\Theta_K) d\Phi_K F^2_{\gamma}(\hat{\mathbf{K}},\hat{\mathbf{k}}) \nonumber\\
&\times& \int \frac{d\Omega_K}{\Omega_K} \, \bigg\{ \Big(\frac{1}{2} + \bar{n}_{pgw} \Big) \big( 4\sin^2(\frac{\Omega_K t}{2}) \big) \nonumber\\
&+&\frac{1}{2} \sinh 2r_K(\eta_H) \, \Re\Big[ e^{2i\phi_K} \big( e^{i\Omega_K t} -1 \big)^2\Big] \bigg\} \,, 
\end{eqnarray}
in which the small-detector approximation $KL \ll 1$ has been assumed. In the above equation, the mean number of gravitons in two-mode squeezed state is defined by $\bar{n}_{pgw} = \sinh^2 r_K$. Hence, the first term represents the contribution of vacuum fluctuations together with the mean number of gravitons in phase diffusion, as if the gravitons were in the thermal state. However, the second term represents the contribution of quantum correlations between gravitons in the two-mode squeezed state, which drastically affects the behavior of the phase uncertainty. More precisely, one may identify different contributions in the phase noise $\mathcal{N}_{pgw}$ as follows:
\begin{eqnarray}\label{eq:29}
\mathcal{N}_{pgw}(t) &\equiv& \mathcal{N}_{vac}(t) + \mathcal{N}_{th}(t) + \mathcal{N}_{corr}(t) \, ,
\end{eqnarray}
where the effect of vacuum fluctuations, thermal gravitons and quantum correlations are specified by
\begin{eqnarray}\label{eq:30}
\hspace{-0.5cm}\mathcal{N}_{vac}(t) &\equiv& \frac{1}{2\pi^2} \left(\frac{\hbar\omega}{E_{Pl}} \right)^2 \sum_{\gamma=+,\times} \int d(\cos\Theta_K) d\Phi_K F^2_{\gamma}(\hat{\mathbf{K}},\hat{\mathbf{k}}) \nonumber \\
&\times& \int \frac{dK}{K} \, \big( 4\sin^2(\frac{\Omega_K t}{2}) \big) \, \\
\hspace{-0.5cm}\mathcal{N}_{th}(t) &\equiv& \frac{1}{\pi^2} \left(\frac{\hbar\omega}{E_{Pl}} \right)^2 \sum_{\gamma=+,\times} \int d(\cos\Theta_K) d\Phi_K F^2_{\gamma}(\hat{\mathbf{K}},\hat{\mathbf{k}}) \nonumber\\
&\times& \int \frac{dK}{K} \, \bar{n}_{pgw} \, \big(4 \sin^2(\frac{\Omega_K t}{2}) \big) \, \nonumber\\
\hspace{-0.5cm}\mathcal{N}_{corr}(t) &\equiv & \frac{1}{2\pi^2} \left(\frac{\hbar\omega}{E_{Pl}} \right)^2 \sum_{\gamma=+,\times} \int d(\cos\Theta_K) d\Phi_K F^2_{\gamma}(\hat{\mathbf{K}},\hat{\mathbf{k}}) \nonumber \\
&\times& \int \frac{dK}{K} \,\sinh 2r_K(\eta_H) \nonumber\\
&\times& \Big( \cos[2\phi_K] - 2\cos[2\phi_K -\Omega_K t] + \cos[2\phi_K - 2\Omega_K t] \Big) \,. \nonumber
\end{eqnarray}
Due to negligibly small coupling strength $\propto (\hbar\omega/E_{pl})^2 \sim 10^{-58}$ for optical frequency $\omega\simeq 10^{15}\,$Hz, vacuum perturbations of spacetime are unlikely to be sensed; hence, we neglect it in the following calculations. However, due to the very high squeezing amplitude $r_K\gg 1$ of the PGWs, the phase noise is drastically amplified, as investigated in the next section.


\subsection{\label{subsec:3.C} Phase noise and phase diffusion}

In order to find an analytical expression for the noise induced by PGWs, given by Eq.~(\ref{eq:30}), we may proceed as follows. For the ultra-low frequency part of the PGWs spectrum, one has a large squeezing amplitude $r_K \gg 1$. This part of the PGW's spectrum leaves the dominant influence on the EM field, and the interaction gets weaker as the frequency increases. Moreover, the squeezing phase $\phi_K$ is vanishing for the super-horizon scales \cite{arani2025revisiting, grishchuk2001relic}, and we may proceed by setting $\phi_K =0$. With these assumptions and neglecting the effect of vacuum fluctuations, one may rewrite $\mathcal{N}_{pgw}$ in Eq.~(\ref{eq:30}) as follows:
\begin{eqnarray}\label{eq:31}
\mathcal{N}_{pgw}(t) &=& \frac{1}{\pi^2} \left(\frac{\hbar\omega}{E_{Pl}} \right)^2 \sum_{\gamma=+,\times} \int d(\cos\Theta_K) d\Phi_K F^2_{\gamma}(\hat{\mathbf{K}},\hat{\mathbf{k}}) \nonumber \\
&\times& \int \frac{dK}{K} \, \Big( \frac{e^{2r_K}}{4} \Big) \Big( 8 \sin^4\big(\frac{\Omega_K t}{2} \big) \Big) \,.
\end{eqnarray}
For typical GW detectors on the Earth or in space, it is safe to assume that the time scale of the experiment $t$ is such that $\Omega_K t \ll 1$. This condition is fulfilled for the major frequency part of the PGW spectrum, which has the prominent effect. Altogether, Eq.~(\ref{eq:31}) can be written in the following compact form 
\begin{eqnarray}\label{eq:32}
\mathcal{N}_{pgw}(t) &\equiv& 4 \Big( \frac{t}{\tau_c} \Big)^4 \, ,
\end{eqnarray}
where the decoherence time-scale induced by PGW is characterized by
\begin{eqnarray}\label{eq:33}
\hspace{-0.5cm}\tau_c &\equiv& \bigg[ \frac{1}{32\pi^2} \left(\frac{\hbar\omega}{E_{Pl}} \right)^2 \sum_{\gamma=+,\times} \int d(\cos\Theta_K) d\Phi_K F^2_{\gamma}(\hat{\mathbf{K}},\hat{\mathbf{k}}) \nonumber \\
&\times& \int_{\Omega_{E}}^{\Omega_1} d\Omega_K \, \Omega_K^3 \, e^{2r_K} \bigg]^{-\frac{1}{4}} \,.
\end{eqnarray}

\noindent By combining Eq.~(\ref{eq:26}) and Eq.~(\ref{eq:31}) we may write
\begin{eqnarray}\label{eq:34}
\Delta^2\phi(z_s, z_r,t) &=& \Delta^2\phi_0 + 4 \,\Big( \frac{t}{\tau_c} \Big)^4 \, .
\end{eqnarray}
In a standard diffusion process, one has $(\Delta\varphi(t))^2 \propto 2 D t$, where $D$ is called the phase diffusion coefficient, and the variance grows linearly with time. However, in the case of two-mode squeezed PGWs, the variance grows quadratically with time. The $t^4$ behavior in Eq.~(\ref{eq:34}) is an exclusive feature of the two-mode squeezed GWs, which recasts to the usual $t^2$ behavior if one considers the vacuum or a thermal heat bath of gravitons.

The growth of $\mathcal{N}_{pgw}(t)$ depends on the related parameters of the PGWs, as already discussed in the literature, such as the tensor-to-scalar ratio $r_{k_0}$, the inflationary parameter $\beta$, the reheating index $\beta_s$, and the reheating temperature $T_{reh}$ (see, for instance, \cite{arani2025revisiting} for a detailed discussion on the related parameters). The upper and lower limits of the integral in Eq.~(\ref{eq:33}) are usually chosen by imposing observational and theoretical considerations (see \cite{tong2013relic,arani2025revisiting}). 
For typical values of the tensor-to-scalar ratio, namely $r_{k_0} \leq 0.032$ as constrained by Planck PR4 \cite{tristram2022improved} and considering $\beta=-2$, $\beta=1$ and $T_{reh}=10^8\,$GeV which are most favored in the literature, the decoherence time scale outcomes $\tau_{c} \simeq 8\times10^{3}\,$s.

Eq.~(\ref{eq:34}) shows that starting from an initial phase fluctuation $\Delta\phi_0$, the uncertainty grows and phase diffusion occurs as a result of interaction with PGWs. One needs an interaction time of around $10^{4}\,$s to observe the vanishing visibility in an interferometer, due to the the phase diffusion of the EM field in the gravitational wave background. Even if this interaction time is for the moment beyond what can be achieved in optical interferometry, it is important to assess the effect of phase diffusion for future ultra-precise measurements, such as ground- and space-based gravitational wave detection with laser interferometers.


\section{\label{sec:5}Conclusions}

In this work, we established a direct connection between the quantum fluctuations of spacetime and the phase diffusion of an electromagnetic field, within the framework of open quantum systems. By applying the Heisenberg-Langevin approach, we showed that the noise correlations of a gravitational wave background—particularly a two-mode squeezed primordial state—govern the stochastic evolution of the EM field's phase. The unique quartic growth of the phase variance over time is a direct manifestation of the squeezed quantum nature of PGWs and marks a clear departure from the linear or quadratic diffusion typical of thermal or vacuum fluctuations. Although the predicted decoherence timescale $\tau_{c} \simeq 10^4\,$seconds remains beyond current interferometric capabilities, our findings highlight an intrinsic quantum limit on phase coherence that must be considered in the design of next-generation gravitational wave detectors. These results offer not only a new channel to probe the quantum origin of cosmological perturbations but also an avenue to assess fundamental quantum-gravitational interactions in precision optical systems.



\bibliography{main}

\end{document}